\documentclass{aa}
\usepackage{natbib}
\usepackage{graphics}
\usepackage{amssymb}
\usepackage{txfonts}
\begin{document}
\title{Molecular freeze-out as a tracer of the thermal and dynamical
evolution of pre- and protostellar cores}

\author{J.K. J{\o}rgensen\inst{1}\thanks{Present address:
Harvard-Smithsonian Center for Astrophysics, 60 Garden St. MS42,
Cambridge, MA 02138, USA} \and F.L. Sch\"{o}ier\inst{2} \and E.F. van
Dishoeck\inst{1}}

\institute{Leiden Observatory, P.O. Box 9513, NL-2300 RA Leiden, The
Netherlands \and Stockholm Observatory, AlbaNova, SE-106 91 Stockholm,
Sweden}
\offprints{Jes K.\,J{\o}rgensen}
\mail{jjorgensen@cfa.harvard.edu} 
\date{Received <date> / Accepted <date>}

\abstract{Radiative transfer models of multi-transition observations
are used to determine molecular abundances as functions of position in
pre- and protostellar cores. The data require a ``drop'' abundance
profile with radius, with high abundances in the outermost regions
probed by low excitation 3~mm lines, and much lower abundances at
intermediate zones probed by higher frequency lines. The results are
illustrated by detailed analysis of CO and HCO$^+$ lines for a subset
of objects. We propose a scenario in which the molecules are frozen
out in a region of the envelope where the temperature is low enough
($\lesssim 40$ K) to prevent immediate desorption, but where the
density is high enough ($>$10$^4$--10$^5$~cm$^{-3}$) that the
freeze-out timescales are shorter than the lifetime of the core. The
size of the freeze-out zone is thereby a record of the thermal and
dynamical evolution of the cores. Fits to CO data for a sample of 16
objects indicate that the size of the freeze-out zone decreases
significantly between Class 0 and I objects, explaining the variations
in, for example, CO abundances with envelope masses. However, the
corresponding timescales are $10^{5\pm 0.5}$~years, with no
significant difference between Class 0 and I objects. These timescales
suggest that the dense pre-stellar phase with heavy depletions lasts
only a short time, of order $10^5$ yr, in agreement with recent
chemical-dynamical models.

} \maketitle

\section{Introduction}
The environments of the youngest pre- and protostellar objects are
characterized by large amounts of cold gas and dust. The chemistry in
these early stages is affected to a large degree by freeze-out of
molecules onto dust grains \citep[e.g.,][]{bergin97} and is closely
related to the thermal evolution of the cores.  An important question
is how the pre- and protostellar stages are linked and what their
respective timescales are. In the pre-stellar stages the thermal
balance is dominated by the external radiation field, which, except in
special cases, does not heat the material to temperatures higher than
$\approx 15$~K \citep[e.g.,][]{evans01}. At such low temperatures most
molecules gradually freeze out, with very short timescales in the
innermost dense regions and increasing timescales toward the exterior
where they may become longer than the age of the core
\citep[e.g.,][]{caselli99}.

In the protostellar stages, in contrast, the thermal balance is
dominated by the heating from the central newly formed protostar,
introducing a steep temperature gradient toward the core
center. Radiative transfer modeling of the dust continuum emission
shows that the characteristic temperatures can rise to a few hundred K
in the innermost regions
\citep[e.g.,][]{shirley02,jorgensen02}. Still, significant depletions
are observed also in these stages, e.g., for CO, indicating that a
substantial fraction of the envelope material remains at low
temperatures \citep[e.g.,][]{blake95,ceccarelli01}. In a large survey
of pre- and protostellar objects, \cite{jorgensen02,paperii} found a
strong correlation between the abundances of CO (and related species
such as HCO$^{+}$) and envelope mass.

The timescales for freeze-out and evaporation depend sensitively on
density and temperature.  The availability of accurate physical
structures from dust continuum data thus provides an opportunity to
constrain the timescales independently using only
chemistry. Currently, the ages of pre- and protostellar objects are
determined almost exclusively from statistics, such as the number
counts of cores with and without associated far-infrared (IRAS)
sources \citep[e.g.,][]{lee99,jessop00}, resulting in a large spread
in pre-stellar ages from $\sim$10$^5$ to a few$\times 10^6$ yr.  Our
semi-empirical procedure for constraining the chemistry and its
assumptions are summarized in Fig.\ 1 of \cite{doty04}, and consists
in the simplest case of fitting a constant abundance to the data, as
used in \cite{jorgensen02,paperii}. In those papers, it was also
realized that constant abundance models provide poor fits to the
lowest excitation lines of species such as CO and HCO$^+$, leading to
the proposal of a ``drop'' abundance profile with a specific
freeze-out region over part of the envelope. Such ``drop''
profiles were also found to best reproduce the interferometer data of
H$_2$CO in two sources \citep{hotcorepaper}. In this paper, we further
explore and quantify the ``drop'' abundance profiles for the full set
of 16 objects and relate the size and location of the depletion zones
to pre- and protostellar evolution. This semi-empirical study forms an
important complement to full chemo-hydrodynamical models which follow
the chemistry in time as the matter collapses to form a central star
\citep[e.g.,][]{rawlings92,lee04}.

\section{Model}
The main assumption in our analysis is that the chemical structure in
the protostellar stages is controlled by thermal desorption
processes. Although other non-thermal processes such as cosmic-ray
induced desorption play a role for weakly-bound species like CO, they
cannot prevent freeze-out in the densest and coldest gas
\citep[e.g.,][]{shen04}. Similarly, photodesorption can keep molecules
off the grains, but this only involves the outermost regions up to an
$A_V\approx 2$ \citep{bergin95}. These effects are expected to be
secondary to the abundance structure resulting from the freeze-out and
thermal evaporation of CO. Finally, outflows and shocks may be
important in regulating the abundance structures, but the narrow
line-widths for optically thin species such as C$^{18}$O and
H$^{13}$CO$^+$ suggest that they probe predominantly the quiescent
bulk envelope material. Our analysis focusses on the chemistry in the
outer envelope on 500--10,000 AU scales, and does not consider the
additional abundance jumps in the innermost region ($<$100 AU) at $T>
100$~K where all ices evaporate.

Following \cite{rodgers03}, the thermal desorption rate $\xi$ and the
freeze-out rate $\lambda$ can be written as:
\begin{eqnarray}
& \xi({\rm M}) & \propto \,\,\exp\left(-\frac{E_{\rm b}({\rm M})}{kT_{\rm d}}\right)  \label{desorpeq} \\ 
& \lambda({\rm M}) & = 4.55\times 10^{-18}\left(\frac{T_{\rm g}}{m({\rm M})}\right)^{0.5}n_{\rm H} \qquad [{\rm s}^{-1}] \label{freezeeq}
\end{eqnarray}
where $T_{\rm d}$ and $T_{\rm g}$ are the dust and gas temperatures,
respectively, $m({\rm M})$ the molecular weight, and $n_{\rm H}$ the
total hydrogen density.  $E_{\rm b}({\rm M})$ is the binding energy of
the molecule depending on the ice mantle composition, for which we
adopt the values tabulated by \cite{aikawa97} ($E_{\rm b}$=960~K for
CO). The CO desorption temperatures of 30~K or higher inferred from
the observational data \citep{jorgensen02} suggest that not all CO is
bound in a pure CO ice matrix but that at least some of it is in a
mixture with H$_2$O where the binding energy may increase to
$\approx$1200~K \citep{collings03apss}. The pre-exponential
factor in Eq.~(\ref{desorpeq}) depends on whether ``zeroth'' or
``first'' order desorption kinetics are considered. In astrochemical
models first-order kinetics are commonly used, which is appropriate
for the desorption of (sub-)monolayer quantities of an adsorbate from
a solid surface. For a thick layer of pure CO, a zeroeth-order
formulation of the desorption process is more appropriate, however
\citep[see discussion in][]{collings03apss}. Our calculations adopt a
first-order formulation with a pre-exponential factor of 10$^{13}$ but
this is not expected to change our main conclusions since the exponent
stays the same and dominates the temperature behavior.

\begin{figure}
\resizebox{\hsize}{!}{\includegraphics{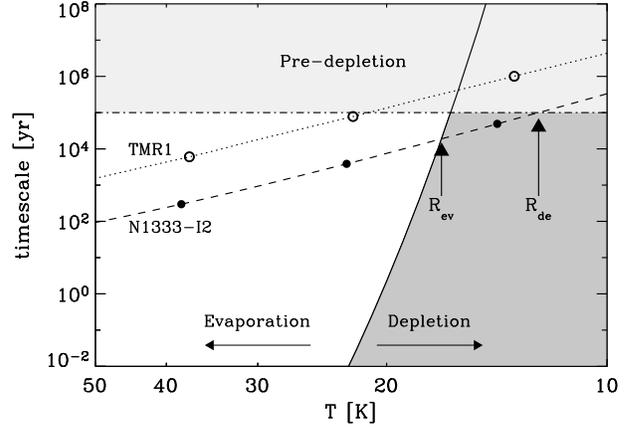}}
\caption{Comparison between CO desorption and freeze-out timescales as
functions of temperature and density. The solid line indicates the
desorption timescale while the dotted and dashed lines indicate the
(density and temperature dependent) freeze-out timescales for
\object{TMR1} and \object{N1333-I2}, respectively. Depletion occurs
where the curves for the freeze-out timescale intersect the dark
colored region, in this example for an assumed age of 10$^{5}$ years
(dash-dotted line). Freeze-out timescales corresponding to H$_2$
densities of 1$\times 10^{5}$, 1$\times 10^{6}$ and 1$\times
10^{7}$~cm$^{-3}$ for NGC~1333-IRAS2 and 1$\times 10^{4}$, 1$\times
10^{5}$ and 1$\times 10^{6}$~cm$^{-3}$ for TMR1 have been indicated by
the filled and open circles, respectively.}\label{timescales}
\end{figure}

Fig.~\ref{timescales} shows the desorption and freeze-out timescales
of CO, defined as $1/\xi({\rm M})$ and $1/\lambda({\rm M})$,
respectively, as functions of depth for two protostars from the sample
of \cite{jorgensen02}, NGC~1333-IRAS2 and TMR1. These objects are
classified as Class 0 and I, and have significantly different envelope
masses of 1.7 and 0.12 M$_{\odot}$, respectively. Their temperatures
and densities have been constrained from submillimeter continuum data
and vary strongly with radius. Figure~\ref{timescales} shows that
freeze-out can -- for a given age -- only occur for a restricted
region of density and temperature in the envelope. At high
temperatures $T>T_{\rm ev}$ the molecule evaporates whereas at low
densities $n<n_{\rm de}$ the freeze-out timescale is too long. Due to
the exponential dependence in Eq.~(\ref{desorpeq}), the thermal
desorption proceeds very rapidly as soon as the temperature is higher
than $T_{\rm ev}$. In contrast, the freeze-out timescale varies more
slowly with depth in the envelope, due to the inverse dependence on
density.

In order to quantify this scenario, a ``drop'' abundance structure is
introduced as a trial profile, with a depleted abundance $X_{\rm D}$
where $T \le T_{\rm ev}$ and $n \ge n_{\rm de}$, and an undepleted
abundance $X_0$ where $T \ge T_{\rm ev}$ or $n \le n_{\rm
de}$. Whereas $T_{\rm ev}$ is in principle a well-defined quantity
depending primarily on the ice mantle properties, $n_{\rm de}$ depends
on the lifetime of the core compared to the depletion timescale
(induced in the pre- and protostellar stages) and the dynamical
evolution of the core at earlier stages, including replenishment of
undepleted material through infall from larger radii. This may move
undepleted envelope material to smaller radii, or higher densities,
increasing the derived value of $n_{\rm de}$. On the other hand,
within the standard inside-out collapse model \citep{shu77} the infall
radii for these very early stages are so small that the bulk of the
envelope is not yet infalling, including the material near the
depletion radius where $n=n_{\rm de}$.

For this discussion we only assume step functions at $T_{\rm ev}$ and
$n_{\rm de}$, which are thus free parameters together with $X_0$ and
$X_{\rm D}$. With these assumptions the full Monte Carlo line
radiative transfer is performed as described in \cite{jorgensen02} and
\cite{schoeier02} using the code of \cite{hogerheijde00vandertak}. We
adopt the molecular data summarized in the \emph{Leiden Atomic and
Molecular Database}
\citep{schoeier03radex}\footnote{\tt{http://www.strw.leidenuniv.nl/$\sim$moldata}}. Single-dish
data on CO and HCO$^+$ taken from \cite{jorgensen02,paperii} are
fitted initially for four objects at different stages of evolution,
leaving all 4 parameters free in the modeling of the CO
data. The abundance of HCO$^+$ is expected to reflect the freeze-out
of CO as illustrated by chemical models
\citep[e.g.,][]{bergin97,lee04}, but the HCO$^+$ lines have
higher critical densities and therefore provide an independent probe.
\citeauthor{lee04} modeled the chemical and physical evolution of a
collapsing core through its pre- and protostellar stages and found
similar ``drop profiles'' for the CO abundances and closely related
species. For the HCO$^+$ fits, $T_{\rm ev}$ and $n_{\rm de}$ are
therefore taken from the fits to the CO data, giving low values of the
$\chi^2$-estimator. With $n_{\rm de}$ constrained from fits to the
observed CO lines, Eq.~(\ref{freezeeq}) then gives the depletion
timescale $t_{\rm de}=1/\lambda$ for a source with a given temperature
and density profile. Subsequently, the CO data for the entire
sample of 16 sources have been fitted, keeping only $X_{\rm D}$ and
$n_{\rm de}$ as free parameters.

\begin{figure}
\resizebox{\hsize}{!}{\includegraphics{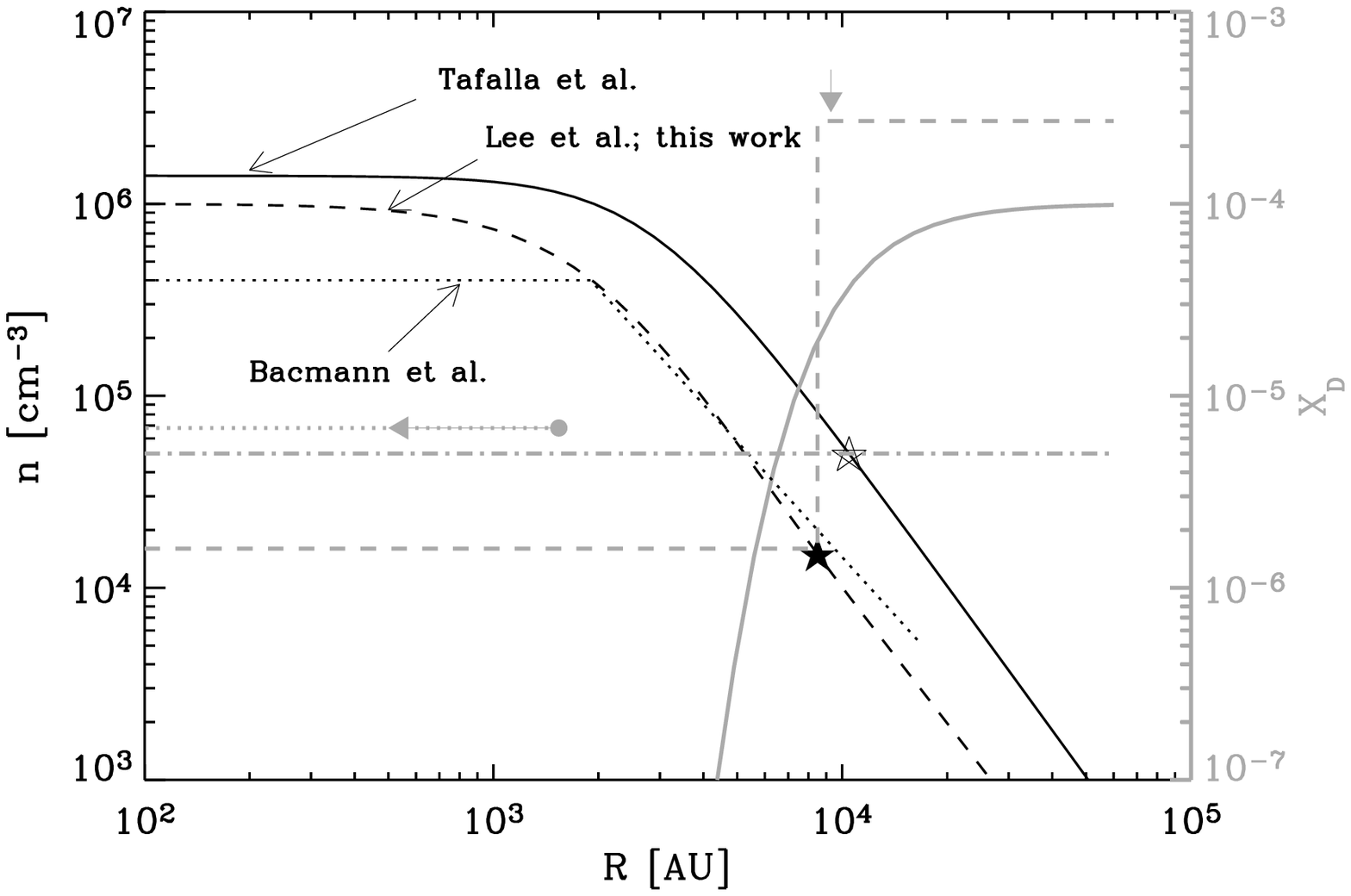}}
\caption{Comparison between the density (black) and CO abundance
profiles (grey) from \cite{bacmann02} (dotted lines), \cite{tafalla02}
(solid lines), \cite{lee03} and this paper (dashed lines) for the
pre-stellar core L1544. The dashed-dotted line indicates the constant
abundance from \cite{jorgensen02}. The freeze-out radius at which a
significant drop in the CO abundance occurs in the step-function model
of \cite{lee03} is indicated by the vertical arrow. The open and
solid stars indicate the values of $n_0$ \citep{tafalla02} and $n_{\rm
de}$ (this work), respectively. See text for details.
}\label{l1544_comparison}
\end{figure}

The pre-stellar cores require a special discussion because they
do not have a central source of heating so that there is only a single
freeze-out radius $R_{\rm de}$ corresponding to density $n_{\rm
de}$. The precise value of $n_{\rm de}$ depends in this case on the
adopted physical structure and abundance profile. As an example of the
uncertainties, our best-fit models for L1544 are compared with other
recent work by \cite{bacmann02}, \cite{tafalla02} and \cite{lee03} in
Fig.~\ref{l1544_comparison}, which shows the best fit physical
structures together with the derived CO abundance structures. There
are some differences in the adopted underlying density profiles and
derived parameters. For example, \cite{tafalla02} use an exponentially
decreasing CO abundance structure toward the center
$X(r)=X_0\exp(-n(r)/n_0)$ with $n_0 = 5.5\times 10^4$~cm$^{-3}$. The
origin of the difference between their $n_0$ and the lower $n_{\rm
de}$ of $1.5\times 10^{4}$~cm$^{-3}$ derived using step functions of
the CO abundance \citep[][this paper]{lee03} stems from (i) the
slightly lower density profile derived by \cite{evans01} and used by
\citeauthor{lee03} and in this paper; (ii) the higher (undepleted) CO
abundance in the outer region of the core from \citeauthor{lee03} and
this paper, and (iii) the fact that CO disappears completely in the
models of \citeauthor{tafalla02} at radii less than $\sim 5000$~AU.
\citeauthor{lee03} test different abundance profile types and find
that a step function with a characteristic density marked with the
vertical arrow in Fig.~\ref{l1544_comparison} provides the best fit.
\cite{bacmann02} derive only an average abundance toward the center of
the core with a density profile with a less dense inner flattened
region, and therefore do not give constraints on $n_{\rm de}$. Taking
this difference in the underlying density profiles into account their
average (or constant) CO abundance of 7$\times 10^{-6}$ is in good
agreement with the constant abundance of 5$\times 10^{-6}$ for L1544
derived by \cite{jorgensen02}.

\section{Results}
The best fit parameters to the CO and HCO$^+$ data for the 4 sources
are listed in Table~\ref{bestfittab} and illustrated in
Fig.~\ref{n1333i2_co_confidence}--\ref{tmr1l1544_co_confidence}. For
the class I object, TMR1, the data are found to be consistent with a
model without a depletion zone. For CO, $X_0$ is found to be
2.7$\times 10^{-4}$ consistent with the direct measurement of the CO
abundance relative to H$_2$ by \cite{lacy94} in a warm cloud in which
CO is not frozen out. This is also the expected value based on the
results of \cite{jorgensen02} who found a maximum constant CO
abundance of $\sim$2$\times 10^{-4}$. Van der Tak et
al.~\citeyearpar{vandertak00} found a similar maximum constant
abundance for a sample of high-mass YSOs. In the context of the drop
abundance model presented in this paper, this constant abundance
applies to sources such as TMR1, where the drop region is so small
that the abundance profile can be assumed to be essentially constant.
\begin{table}
\caption{Best fit parameters for CO and HCO$^+$ lines for the
protostars N1333-I2, L723, TMR1 and the pre-stellar core L1544.}
\label{bestfittab} \begin{center}
\begin{tabular}{lllll} \hline\hline
                         &  N1333-I2     & L723         & TMR1         & L1544 \\ \hline
$T_{\rm ev}$ [K]         &  $\gtrsim~35$ & 40           & $\ldots$     & $\ldots$ \\
$n_{\rm de}$ [$10^4$~cm$^{-3}$] &  7     & 4            & $\ldots$     & 1.5 \\[2.0ex]
\multicolumn{5}{l}{CO:} \\
$X_0$ [$10^{-4}$]        &  2.7          & 2.7          & 2.7          & 2.7 \\
$X_{\rm D}$ [$10^{-4}$]  &$<$0.14        & 0.14         & $\ldots$     & 0.02 \\
$\chi^2$ / $N$           &  4.4 / 7      & 2.4 / 6      & 0.75 / 6     & 4.2 / 4 \\[2.0ex]
\multicolumn{5}{l}{HCO$^+$:} \\
$X_0$ [$10^{-8}$]        &  1.8          &$<$2.0        & 2.7$^{b}$    & -- \\
$X_{\rm D}$ [$10^{-8}$]  &  0.26         & 0.35         & $\ldots$     & -- \\
$\chi^2$ / $N$           &  4.1 / 5      & 1.8 / 3      & 2.1$^b$ / 2  & -- \\[2.0ex]
$t_{\rm de}$ [$10^5$~yrs]$^{a}$ &  2     & 3            & $\lesssim$~1$^{c}$ & 8 \\  \hline
\end{tabular}\end{center}

Notes: The abundances have been derived from observations of the
optically thin isotopic species, C$^{18}$O, C$^{17}$O and H$^{13}$CO$^+$ assuming
the standard isotopic ratios adopted in
\cite{paperii}. $^{a}$Depletion timescale corresponding to the derived
$n_{\rm de}$ using Eq.~(\ref{freezeeq}). $^{b}$Model including
excitation through collisions with electrons at densities $\le 3\times
10^4$~cm$^{-3}$. $^{c}$1$\sigma$ limit assuming $T_{\rm ev}$ of 35~K.
\end{table}

\begin{figure}
\resizebox{\hsize}{!}{\rotatebox{90}{\includegraphics{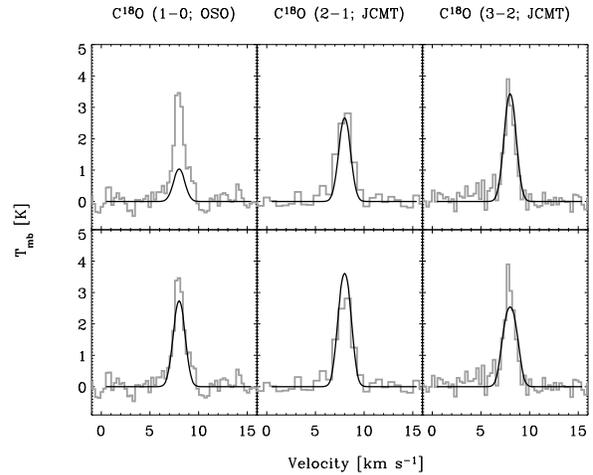}}}
\caption{Fitted C$^{18}$O line-profiles for NGC~1333-IRAS2. Upper
panels: constant fractional abundance [$^{12}$CO] of 2.4$\times
10^{-5}$ from \cite{jorgensen02}. Lower panel: best fit drop abundance
model with $X_0=2.7\times 10^{-4}$ and $X_{\rm D}=1.4\times 10^{-5}$
from this paper. The calibrational uncertainty for each observed
spectrum is about 20\%.}\label{n1333i2_c18o_drop}
\end{figure}

\begin{figure}
\resizebox{\hsize}{!}{\includegraphics{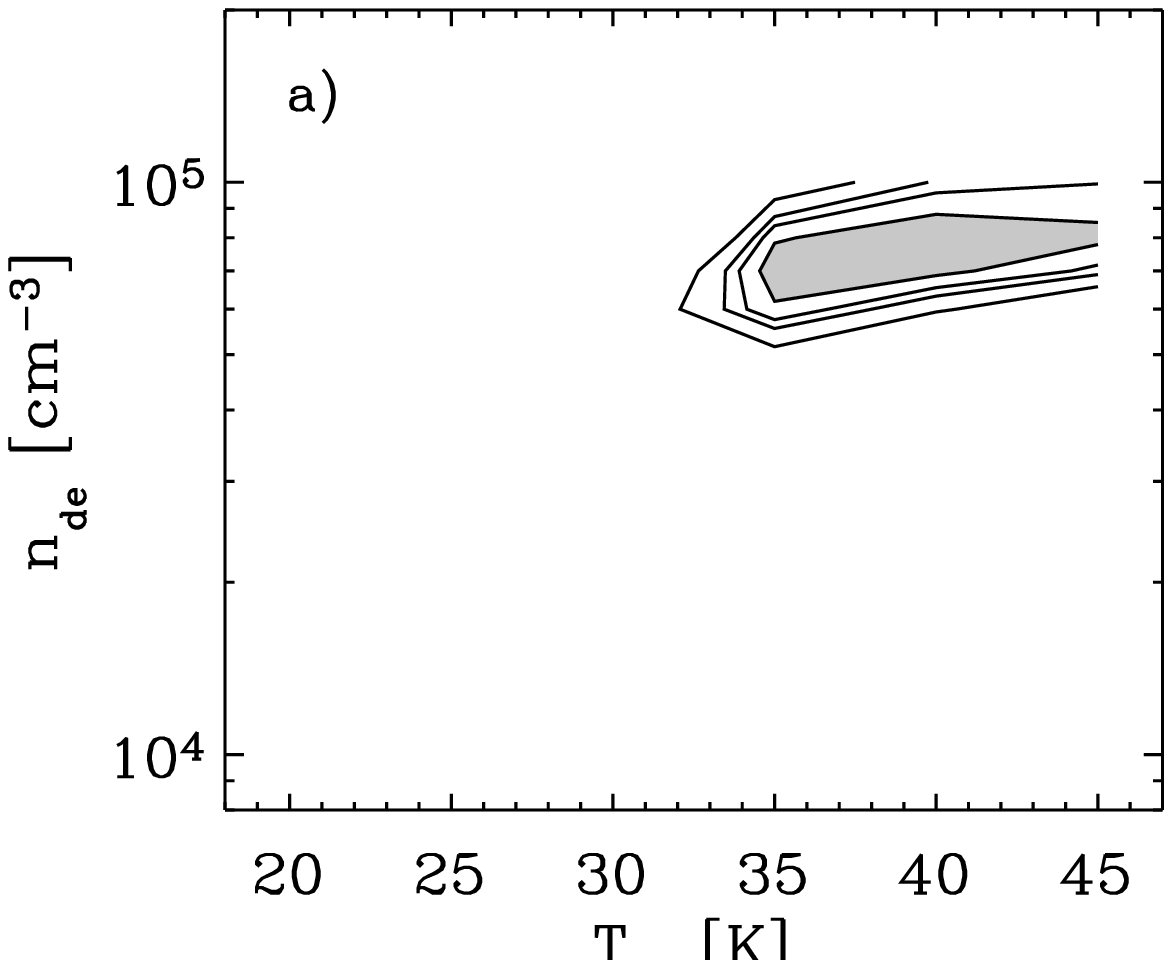}\includegraphics{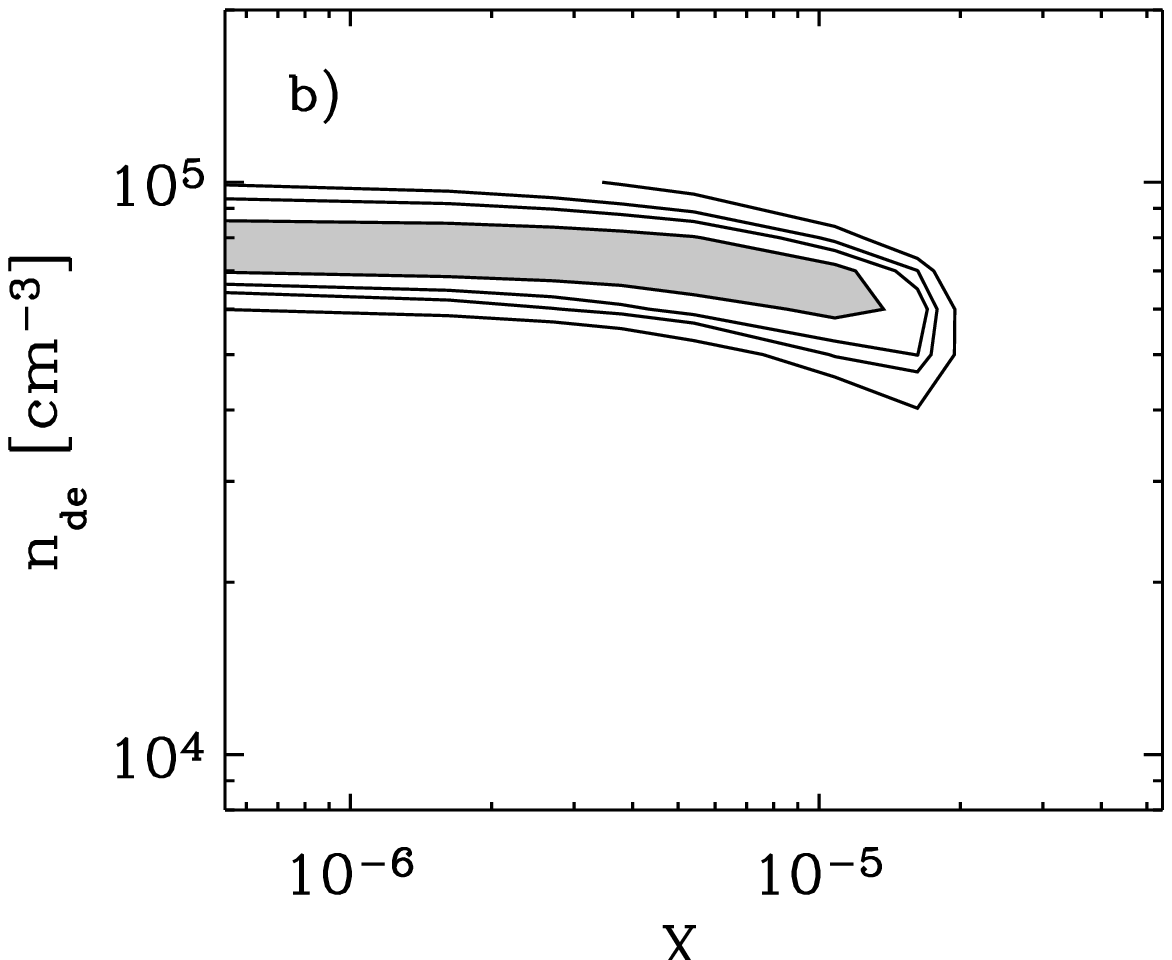}}
\resizebox{\hsize}{!}{\includegraphics{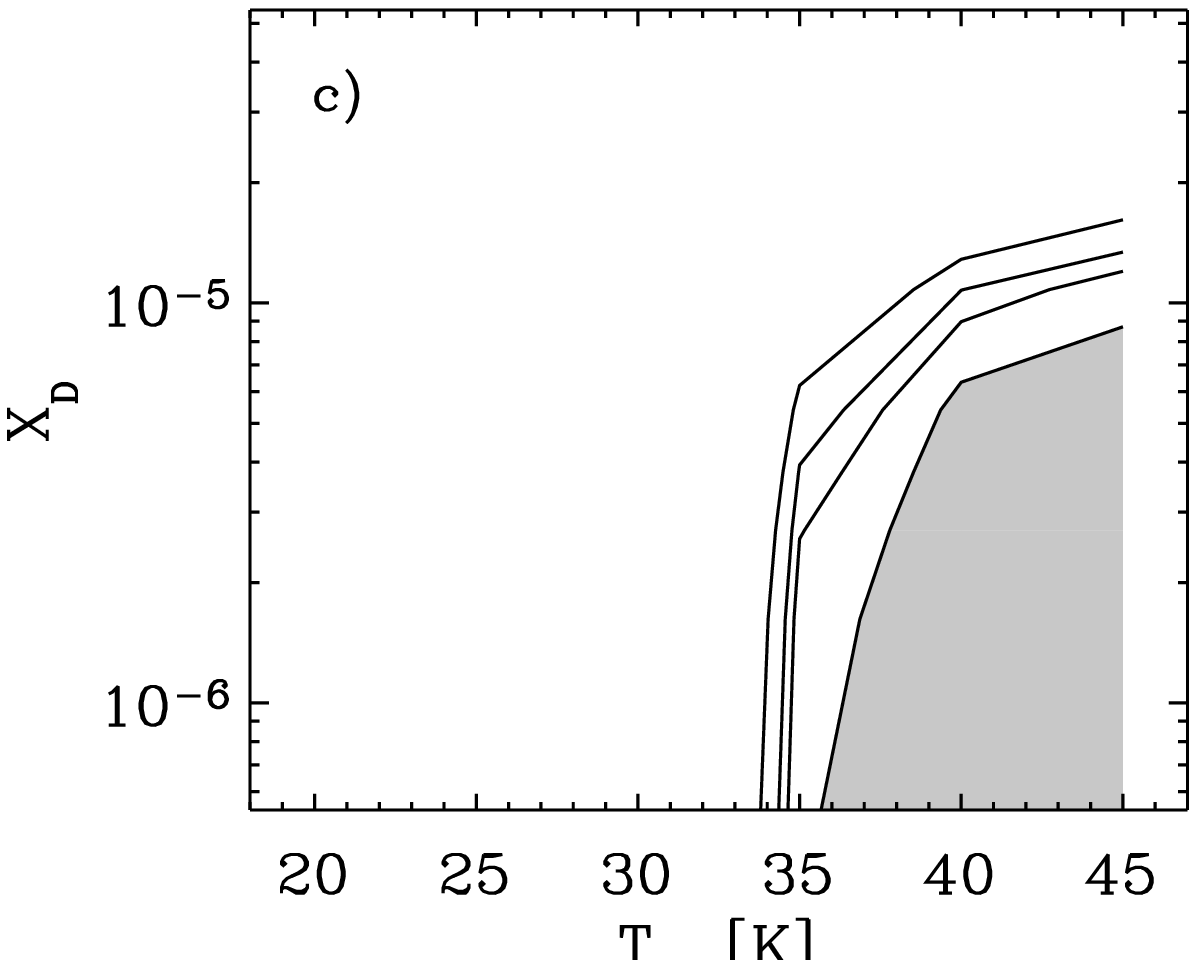}\includegraphics{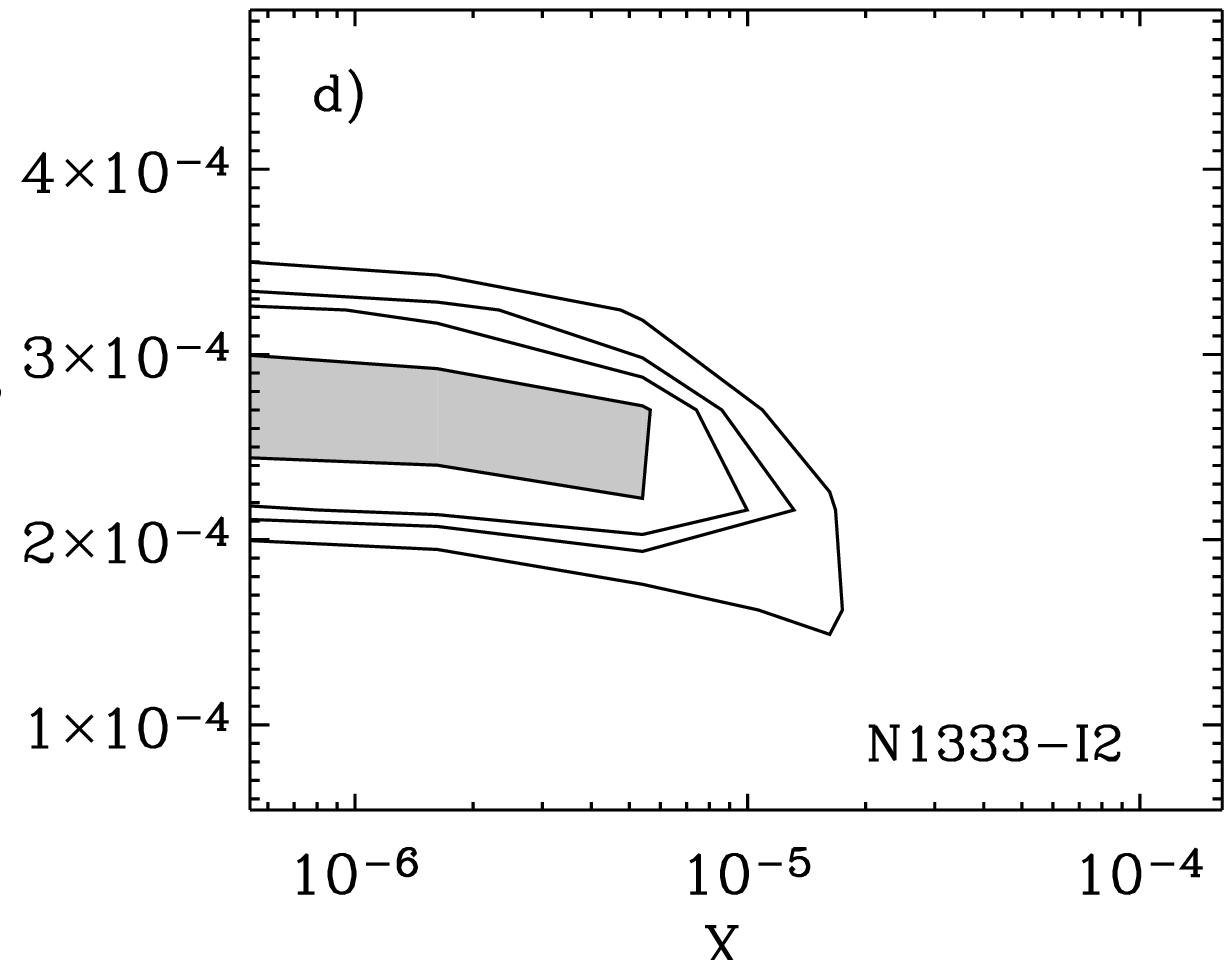}}
\caption{$\chi^2$ confidence plots for fits to the CO lines toward
NGC~1333-IRAS2. In each plot the grey area marks the 1$\sigma$
confidence region and the subsequent line contours the 2$\sigma$,
3$\sigma$ and 4$\sigma$ confidence
regions.}\label{n1333i2_co_confidence}
\end{figure}

\begin{figure}
\resizebox{\hsize}{!}{\includegraphics{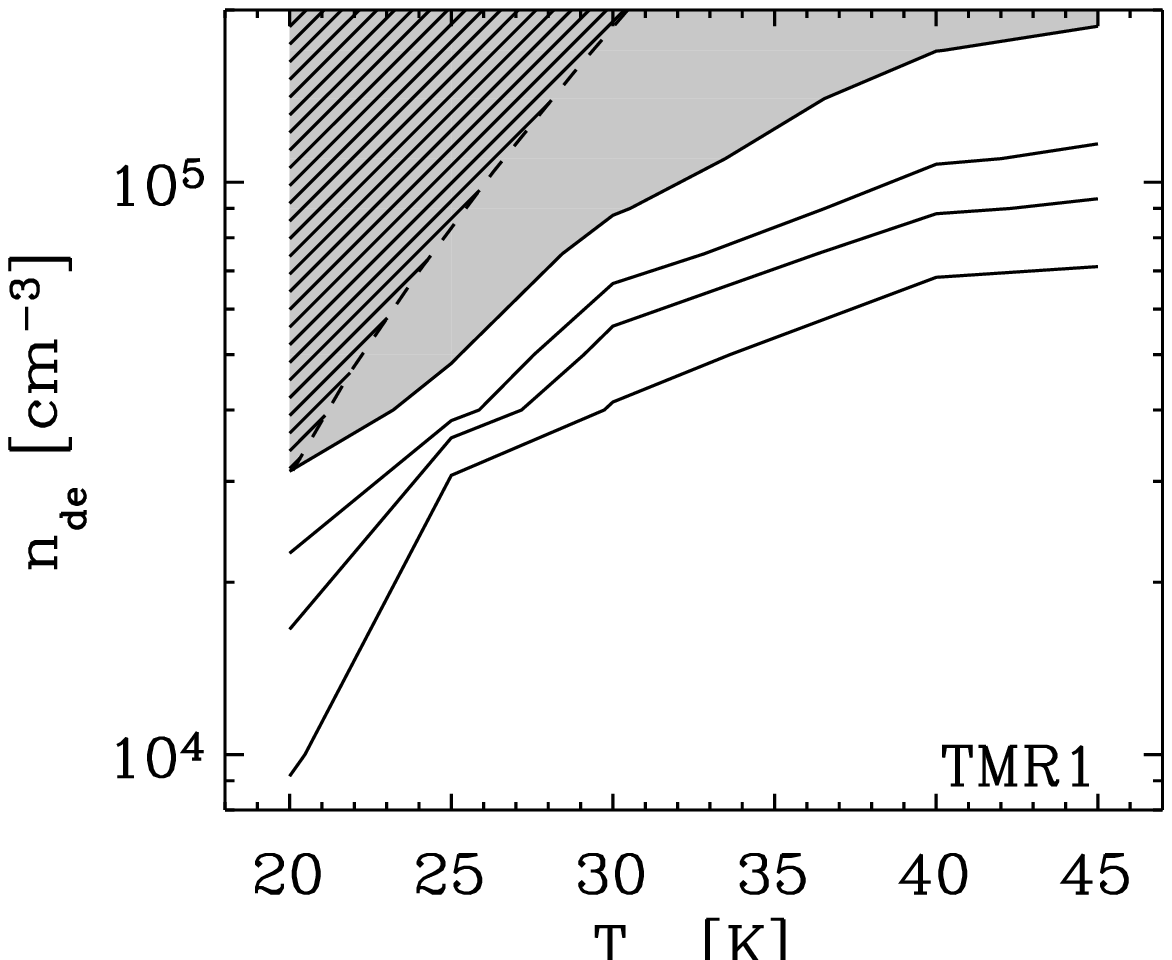}\includegraphics{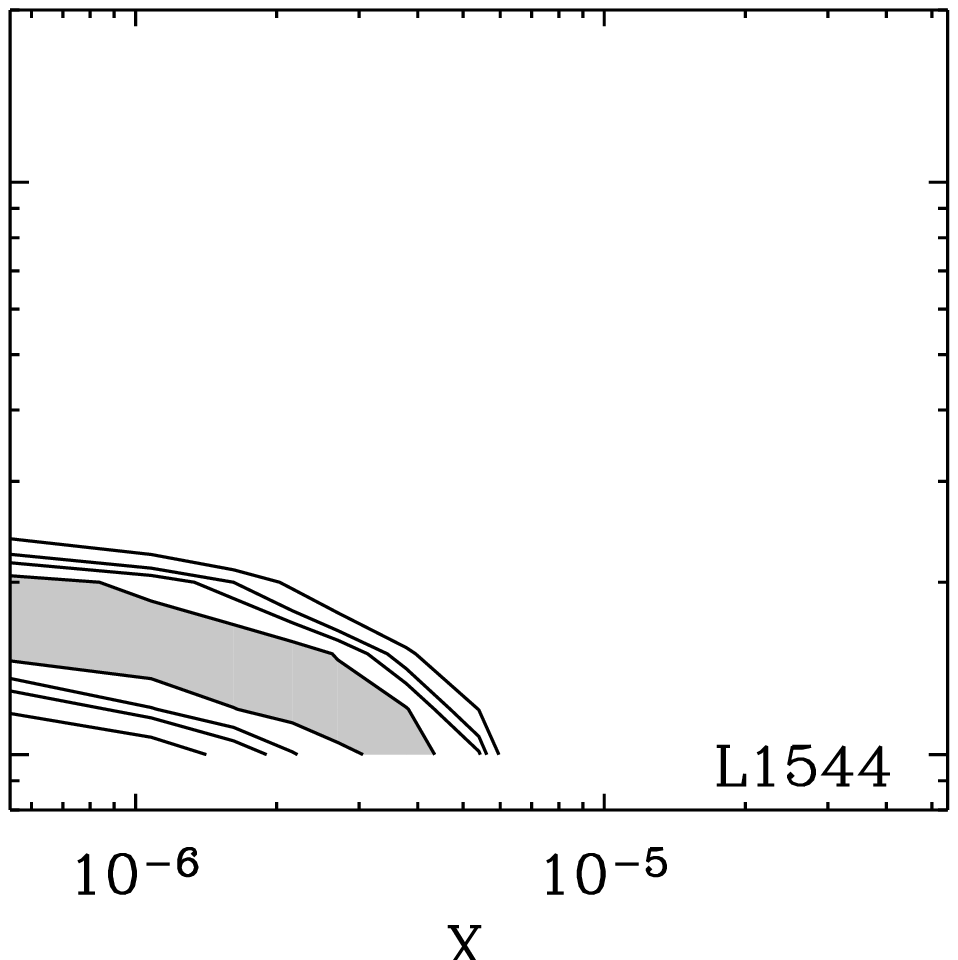}}
\caption{As Fig.~\ref{n1333i2_co_confidence} but for TMR1 (left)
and L1544 (right). In the left panel the dashed region indicates
constant abundance models (i.e., models where the radius corresponding
to $n_{\rm de}$ is located closer to the central source than the radius
corresponding to $T_{\rm ev}$).}\label{tmr1l1544_co_confidence}
\end{figure}

For the class 0 objects, in contrast, the constant abundances found by
\cite{jorgensen02} are about an order of magnitude lower, indicating
that the region of CO depletion (the drop zone) is large for these
objects. These are also the objects where the intensities of the low
excitation $J=1-0$ transitions are underestimated unless the drop
abundance structure is introduced. This is clearly illustrated in
Fig.~\ref{n1333i2_c18o_drop} which shows the best fit to the C$^{18}$O
lines for NGC~1333-IRAS2 with a constant abundance model from
\cite{jorgensen02} and the drop model from this paper. A similar plot
illustrating the fits for L723 can be found in Fig.~7 of
\cite{paperii}. Fig.~\ref{n1333i2_co_confidence} illustrates the
confidence levels for the various parameters for NGC~1333-IRAS2. It is
seen that $T_{\rm ev}$ only has a lower limit and $X_{\rm D}$ only an
upper limit: for $T_{\rm ev}$ higher than 35--40~K the innermost
region with high CO abundances has a very low filling factor in the
single-dish beam and therefore only contributes a small fraction to
the beam-averaged column densities. The best fit value of $X_{\rm 0}$
is again 2.5--3$\times 10^{-4}$.

For all sources listed in Table~\ref{bestfittab} except TMR1,
$X_{\rm D}$ is found to be typically an order of magnitude lower than
$X_0$ for both CO and HCO$^+$. The CO evaporation temperatures
$T_{\rm ev}$ are all found to be $\ge 35$~K, consistent with the
results of \cite{jorgensen02}, whereas the values of $X_{\rm 0}$
are all of order 2--3$\times 10^{-4}$ with a 10--20\% statistical
uncertainty.

The derived CO abundance profile depends somewhat on the adopted outer
radius and any contribution from the surrounding cloud to the lowest
excitation lines. However, the same drop abundance profiles are also
necessary to reproduce higher angular resolution interferometer data,
where contributions from the larger scale cloud are resolved out, as
well as to model transitions with higher critical densities of H$_2$CO
and HCO$^+$ which are not excited in the surrounding cloud
\citep[][and Table~\ref{bestfittab}]{hotcorepaper,l483art}.

Given the success of these models, the C$^{18}$O and C$^{17}$O lines
for the entire sample of \cite{jorgensen02} have been fitted to infer
the CO abundance structure and the depletion time scales. Based on the
results for the above 4 sources, it seems reasonable to take the
undepleted abundance $X_0$ fixed at $2.7\times 10^{-4}$ and the
evaporation temperature $T_{\rm ev}$ at 35~K to reduce the number of
free parameters. We stress that uncertainties in the underlying
physical model such as due to the uncertain dust properties \citep[see
discussion in][]{jorgensen02} and other factors may result in
systematic uncertainties by factors of 2--3 in the derived absolute
abundances and values of $n_{\rm de}$ and $t_{\rm de}$ (see also above
example on L1544). However, the conclusion that a drop abundance
profile provides a much better fit than a constant abundance profile
is not affected by these uncertainties. Also, since all sources are
analyzed in the same way, the relative values and inferred trends
should be more reliable.

The fitted $n_{\rm de}$ and $X_{\rm D}$, together with the $\chi^2$
value and derived $t_{\rm de}$ are given in
Table~\ref{sample_fits}. For each source, all lines could be well
fitted ($\chi^2_{\rm red} \le 2$) using depletion densities $n_{\rm
de}=$1$\times 10^{4}$--6$\times 10^{5}$~cm$^{-3}$ ($10^{5\pm
0.4}$~cm$^{-3}$) corresponding to depletion timescales of $2\times
10^4$--$8\times 10^5$~years ($10^{5\pm 0.5}$~years). No significant
difference in timescales between Class 0 and I objects is found. For
the objects with the least massive envelopes ($M\lesssim 0.1~M_\odot$)
where photodesorption could become an issue, the derived $t_{\rm de}$
is a lower limit to the actual depletion timescale. In general,
however, the depletion radius is located further in the envelope than
the radius where $A_{\rm V}=2$ for our objects.


\begin{table}
\caption{Values of $n_{\rm d}$ and $X_{\rm D}$ constrained from the CO
lines for each individual source in the sample, together with the
total $\chi^2$, number of fitted lines and derived $t_{\rm
de}$.}\label{sample_fits} \begin{center}
\begin{tabular}{lllll}\hline\hline
Source    & $n_{\rm de}$  & $X_{\rm D}$ & $\chi^2$ / $N$ & $t_{\rm de}$ \\ 
          & [cm$^{-3}$]   &             &                &  [yr]        \\ \hline
L1448-I2  &  4.8$\times 10^{5}$   & 3.0$\times 10^{-7}$ & 4.2 / 4        & 3$\times 10^{4}$     \\
L1448-C   &  6.0$\times 10^{4}$   & 2.0$\times 10^{-5}$ & 10.7 / 7       & 2$\times 10^{5}$     \\
N1333-I2  &  7.0$\times 10^{4}$   & 1.4$\times 10^{-5}$ & 4.4 / 7        & 2$\times 10^{5}$     \\
N1333-I4A &  6.0$\times 10^{5}$   & 2.0$\times 10^{-7}$ & 11.2 / 7       & 2$\times 10^{4}$     \\
N1333-I4B &  1.7$\times 10^{5}$   & 4.3$\times 10^{-6}$ & 9.5 / 7        & 7$\times 10^{4}$     \\
L1527     &  1.4$\times 10^{5}$   & 2.7$\times 10^{-5}$ & 9.2 / 7        & 1$\times 10^{5}$     \\
VLA1623$^{a}$ &  --           & --          & --             & --           \\
L483$^b$  &  1.5$\times 10^{5}$   & 5.0$\times 10^{-6}$ & 1.5 / 3        & 9$\times 10^{4}$     \\
L723      &  4.0$\times 10^{4}$   & 1.4$\times 10^{-5}$ & 2.4 / 6        & 3$\times 10^{5}$     \\
L1157     &  $<$2.3$\times 10^{5}$& --          & --             & $>$5$\times 10^{4}$  \\
CB244     &  1.3$\times 10^{5}$   & 1.6$\times 10^{-5}$ & 3.8 / 5        & 1$\times 10^{5}$     \\
L1489     &  7.0$\times 10^{4}$   & 1.0$\times 10^{-5}$ & 8.1 / 6        & 1$\times 10^{5}$     \\
TMR1      & $\gtrsim$1.0$\times 10^{5}$ & --    & 0.75 / 6       & $\lesssim$8$\times 10^{4}$ \\
L1544     &  1.5$\times 10^{4}$   & 1.6$\times 10^{-6}$ & 4.2 / 4        & 8$\times 10^{5}$     \\
L1689B    & $\gtrsim$1.0$\times 10^{4}$ & $\lesssim$5$\times 10^{-5}$ & 0.16 / 3 & $\lesssim$1$\times 10^{6}$     \\ 
IRAS16293$^{c}$ &  1.0$\times 10^{5}$   & --    & --             & 1$\times 10^{5}$ \\ \hline
\end{tabular}\end{center}

$^{a}$For VLA1623 the 1--0 line observations are significantly
underestimated even for a high constant abundance. This likely
reflects the dense ridge of material in which this source is located, which
also affects the higher excitation lines \cite[see also discussion
in][]{jorgensen02}. $^{b}$From fits to single-dish and interferometric
C$^{18}$O observations \citep{l483art}. $^{c}$From fits to H$_2$CO
interferometer data \citep{hotcorepaper}. The CO isotopic lines for
IRAS~16293-2422 are consistent with a constant abundance throughout
the envelope but do not rule out a drop in abundance.
\end{table}

\section{Discussion}\label{discuss}
\begin{figure}
\resizebox{\hsize}{!}{\includegraphics{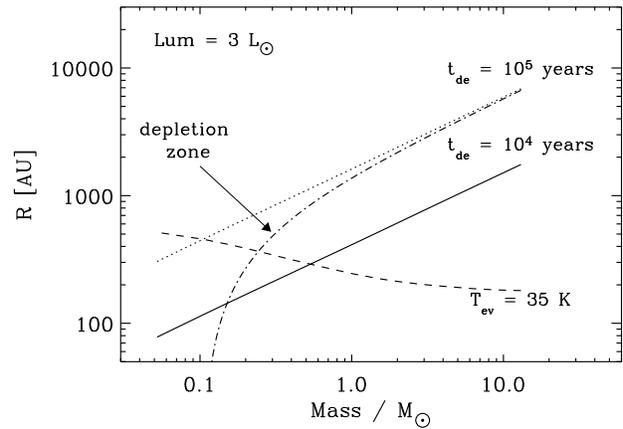}}
\caption{Characteristic radii as functions of envelope mass for a
3~$L_\odot$ protostar using the models of \cite{jorgensen02}. An
$n\propto r^{-1.5}$ envelope density profile was assumed with inner
and outer radii of 25 and 15000~AU, respectively. The dashed line
indicate the radius where $T_{\rm ev} = 35$~K and the solid and dotted
lines the radii where the depletion timescales are 10$^{4}$ and
10$^{5}$~years, respectively. The dashed-dotted line indicates the
difference between the depletion and evaporation radii (i.e., size of
the depletion zone) for a depletion timescale of 10$^{5}$
years.}\label{time_evol}
\end{figure}

The fact that the drop abundance profiles provide better fits than
constant abundance profiles for all sources and molecules studied
shows that this structure provides a good representation of the
dominant chemistry in protostellar envelopes, with other chemical
effects being of secondary importance. To illustrate the dependence on
envelope mass, Fig.~\ref{time_evol} compares the radii corresponding
to $T_{\rm ev}$ and $n_{\rm de}$ for varying envelope masses for a
(typical) 3~$L_\odot$ central source. For an object with a large
envelope mass like NGC~1333-IRAS2 (1.7~$M_\odot$), the depletion zone
is a few thousand AU, a substantial fraction of the entire
envelope. For an object with a small envelope mass like TMR1
(0.1~$M_\odot$), the freeze-out radius moves inwards and the size of
the depleted region becomes vanishingly small. This indicates that the
trend of increasing (constant) abundances with decreasing envelope
mass for these species \citep{jorgensen02,paperii} reflects the size
of the depletion region.

\begin{figure}
\begin{center}
\resizebox{\hsize}{!}{\includegraphics{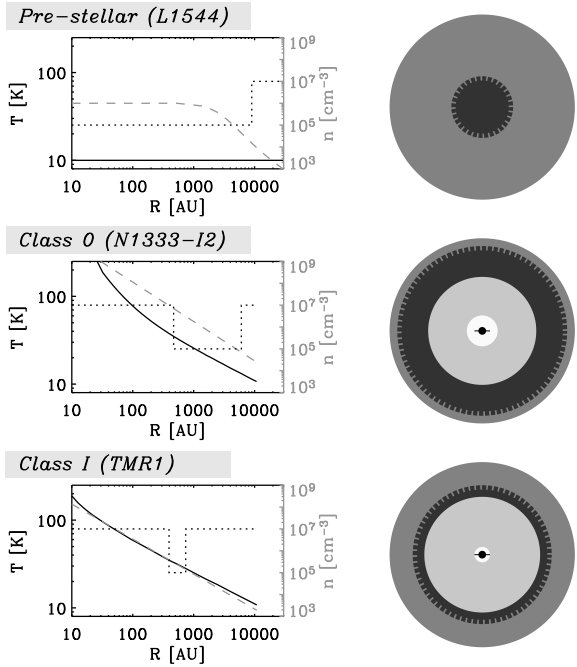}}
\end{center}
\caption{Proposed chemical structure for low-mass pre- and
protostellar objects. The left column gives the temperature and
density as functions of radius (black solid and grey dashed lines,
respectively) for three archetypical low-mass pre- and protostellar
objects: \object{L1544} (pre-stellar core), \object{N1333-I2} (class
0, $M_{\rm env}>0.5$ M$_{\odot}$ protostar) and \object{TMR1} (class
I, $M_{\rm env}< 0.5$ M$_{\odot}$ protostar). The black dotted lines
indicate the derived abundance structures. The right column gives the
depletion signature for each class of object with, going from the
outside to the inside, the dark grey indicating the region where the
density is too low for depletion ($n< n_{\rm de}$), the black
indicating the region where the molecules deplete and the light grey
indicating the region where they evaporate ($T > T_{\rm
ev}$).}\label{chemicalscenario}
\end{figure}

Based on the above fits, we propose a chemical structure of the outer
envelopes as shown schematically in Fig.~\ref{chemicalscenario}. The
main difference between the pre- and protostellar cores is the inner
source of heating in the protostars that causes CO to be evaporated
rather than depleted toward the source center. The main difference
between the Class 0 and I objects is the size of the depletion zone.

Is this chemical structure also an evolutionary indicator? Due to
progressive dispersion of the envelope by outflows and mass accretion,
an intially massive envelope like that of NGC~1333-IRAS2 should
eventually go through a phase with a low-mass envelope like that
around TMR1. However, in contrast with the size of the depletion
zone, no clear correlation is seen between the derived age $t_{\rm
de}$ and derived envelope mass or luminosity for the entire sample
(e.g., Table~1). This indicates that the depletion timescales must
also reflect other properties, such as the mass of the core from which
the protostar is formed. In other words, an other object like
TMR1 did not necessarily start out with an envelope as massive as that
of NGC~1333-IRAS2. The relatively short timescales of $10^{5\pm
0.5}$~years suggest that the depletion structure is established in the
pre-stellar stages, but only after the pre-stellar core has become
dense enough that freeze-out really sets in. The rather small scatter
seen indicates that this stage is short, of order $10^5$~years.

This timescale provides an interesting independent age constraint for
the studies of core collapse, since it is shorter than the ages
derived from the above mentioned statistical studies
\citep[e.g.,][]{lee99,jessop00}. This apparent discrepancy may reflect
simply the definition of the pre-stellar stage: our estimate refers
only to the dense pre-stellar phase where depletion has become
significant. On the other hand, the statistical studies may be missing
low-luminosity embedded infrared sources due to the limited sensitivty
of IRAS \citep[see, e.g.,][]{young04} and thus lead to overestimates
of the timescales. Further unbiased surveys, e.g., with the Spitzer
Space Telescope, may shed further light into this issue.

To refine the proposed models, it will be necessary to couple models
for the dynamical and radial chemical evolution as done by
\cite{lee04} to fully address the use of $n_{\rm de}$ as a tracer of
age. It is interesting to note that our derived empirical drop
abundance structure agrees well with these detailed models, not only
qualitatively but also quantitatively. As their Fig.~6 shows, the
timescales over which the heavy depletion occurs in the pre- and
protostellar stages is only of order $10^5$~years.

Another important consequence of the ``drop'' chemical structure is
that it affects the tracers of infall in protostostellar
envelopes. Since the dynamical structure of protostellar cores is
often inferred from fits of line profiles of molecules such as HCO$^+$
\citep[e.g.,][]{gregersen97}, knowledge about the radial chemical
structure is important for detailed descriptions of the infalling
envelope. For example, the location of the collapse radius in the
inside-out collapse model for the envelope around NGC~1333-IRAS2
\citep{n1333i2art} is at $\approx 1000$~AU which is in the middle of
the drop zone where the temperature is $\approx 25$~K. The exact
``infall'' line profile will therefore depend critically not only on
the velocity field but also the presence and location of the outer
pre-depletion ($n < n_{\rm de}$) zone and the amount of depletion.

\begin{acknowledgement}
The authors thank Ted Bergin, Jeong-Eun Lee and Neal Evans for useful
discussions. The research of JKJ is made possible through a NOVA
network 2 Ph.D. stipend, FLS acknowledges support from the Swedish
Research Council. Astrochemistry research in Leiden is supported by a
NWO Spinoza grant.
\end{acknowledgement}

\end{document}